\documentclass{aa}
\usepackage{graphicx}
\usepackage{txfonts}

\begin{document}

\title{The pre- versus post-main sequence evolutionary phase of B[e] stars}
\subtitle{Constraints from $^{13}$CO band emission}

\author{M. Kraus}

\institute{Astronomick\'y \'ustav, Akademie v\v{e}d \v{C}esk\'e republiky, Fri\v{c}ova 298, 251~65 Ond\v{r}ejov, Czech Republic\\
           \email{kraus@sunstel.asu.cas.cz}
      }

\date{Received; accepted}

\abstract
{Many galactic B[e] stars suffer from improper distance determinations, 
which make it difficult to distinguish between a pre- and post-main sequence 
evolutionary phase on the basis of luminosity arguments. In addition, these
stars have opaque circumstellar material, obscuring the central 
star, so that no detailed surface abundance studies can be performed.}
{Instead of studying the surface abundances as a tracer of the evolutionary
phase, we propose a different indicator for the supergiant status of a B[e] 
star, based on the enrichment of its circumstellar matter by $^{13}$C, and 
detectable via its $^{13}$CO band emission in the $K$ band spectra.}
{Based on stellar evolution models, we calculate the variation of the 
$^{12}$C/$^{13}$C isotopic surface abundance ratio during the evolution of
non-rotating stars with different initial masses. For different values of the
$^{12}$C/$^{13}$C ratio we then compute synthetic first-overtone 
vibration-rotational band spectra from both the $^{12}$CO and $^{13}$CO 
molecule at different spectral resolutions. We further discuss the influence of 
stellar rotation on the variation of the surface $^{12}$C/$^{13}$C ratio and on 
the possibility of $^{13}$CO band detection.}
{The surface $^{12}$C/$^{13}$C isotope ratio is found to decrease strongly  
during the post-main sequence evolution of non-rotating stars, from its 
interstellar value of about 70 to a value of about 15--20 for stars with 
initial masses higher than 7\,M$_{\odot}$, and to a value of less than 5 for 
stars with initial masses higher than 25\,M$_{\odot}$. We find that detectable 
$^{13}$CO band head emission is produced for isotope ratios $^{12}$C/$^{13}{\rm 
C}\la 20$, and can most easily be detected with a spectral resolution 
of $R\sim 1500 \ldots 3000$. For the rotating stellar models, the drop in 
$^{12}$C/$^{13}$C already occurs for all stars with $M_{\rm in} \ga 
9\,M_{\odot}$ during the main-sequence evolution. The detection of 
$^{13}$CO band head emission in such mid-resolution $K$ band spectra of a B[e]
star thus favours an evolved rather than a young nature of the object.} 
{}

\keywords{Stars: early-type -- stars: atmospheres -- stars: mass-loss -- stars: winds, outflows -- circumstellar matter}

\maketitle

\section{Introduction}\label{intro}

The classification of B[e] stars according to their evolutionary phase is a
long-standing problem. Much effort has been undertaken since Lamers et al.
(\cite{Lamers98}) published the first detailed classification criteria
sorting the B[e] stars into the classes of supergiants, Herbig stars, compact
planetary nebulae, and symbiotics. Nevertheless, the remaining group of
unclassified B[e] stars, which includes about half of all known B[e] stars, has
since then gained even more members.
                                                                                
While many targets within the group of unclassified B[e] stars have  not
been studied yet in detail, there also exists a number of stars showing
indications of both a young (Herbig star) as well as an evolved (supergiant
star) nature. And one of the reasons for their uncertain classification is 
their still rather poorly known distance, and hence luminosity (see
Table\,\ref{tab_sg}). For a proper
assignment of an evolutionary phase to these stars it is, therefore,
necessary to find other characteristics that are directly linked to the
evolutionary phase of a star as additional classification criteria.

The best studied B[e] supergiant sample is located in the Magellanic Clouds.
Their main characteristic is the so-called hybrid character of their spectra,
showing indications for both a typical line-driven wind in polar direction,
and a slow and low-ionized but high-density wind in the equatorial direction
(Zickgraf et al.\,\cite{Zick85}). Further evidence for non-sphericity of their
winds comes from polarimetric observations, e.g., by Magalh\~aes
(\cite{Magalhaes}), Magalh\~aes et al. (\cite{Magalhaesetal}), and Melgarejo et
al. (\cite{Melgarejo}) for the Magellanic Cloud stars, and from, e.g., Zickgraf 
\& Schulte-Ladbeck (\cite{ZickSchulte}) for some of the galactic candidates.
Detection of molecular emission bands of TiO (Zickgraf et al.\,\cite{Zick89})
and CO (McGregor et al. \cite{McGregor, McGregor89}; Morris et al. 
\cite{Morris}) in the spectra of several B[e] supergiants confirms the presence 
of high-density and cool material in the vicinity of these luminous stars.

An additional characteristic of B[e] supergiants is their strong infrared (IR) 
excess emission (e.g., Zickgraf et al.\,\cite{Zick86}), indicating that these 
luminous objects are surrounded by hot or warm dust, whose most probable 
location is within a massive circumstellar disk. Recent estimates, based on 
observations with the {\it Spitzer Space Telescope} Infrared Spectrograph for
the Large Magellanic Cloud B[e] supergiant R\,126 by Kastner et al.
(\cite{Kastner}), resulted in a total dust mass within the disk of $\sim 3
\times 10^{-3}$\,M$_{\odot}$. 

The existence of dusty disks is also one of the major characteristics of Herbig
Ae/Be stars (see, e.g., the recent reviews by Waters \cite{Rens}, and Waters \&
Waelkens \cite{Rens2}, and references therein). And for many young stellar
objects the observations of the CO band heads (Hanson et al. \cite{Hanson}; Bik
\& Thi \cite{BikThi}; Thi et al. \cite{Thi}; Bik et al. \cite{Bik}), although
not present in all objects, often resulted in detailed information about the
disk structure and kinematics of the inner disk. A high-density
circumstellar disk with molecules and dust is thus a common feature of both
the Herbig stars and the B[e] supergiants. The only, and consequently most
important difference is the fact that for the Herbig stars the disk material
is the remnant of the star formation process.

Since the progenitors of B[e] supergiants must have been massive O-type 
stars, this close-by dust cannot be pre-main sequence in origin. Any 
remaining pre-main sequence circumstellar material 
will have been swept away by the strong radiation pressure of their stellar 
winds during their main-sequence life-time. The disks around the B[e] 
supergiants must thus have formed from the wind material itself\footnote{This 
argument is only strictly valid if we assume that the B[e] phenomenon is a 
special phase (for stars with perhaps very specific initial conditions) during 
single star evolution, which we will assume throughout this paper. If the B[e] 
phenomenon for these supergiants is caused by some binary interaction or binary 
merger process, the situation might be completely different.}, due to some 
enhanced mass loss or some violent high-density mass ejection, predominantly 
within the equatorial regions. Detailed investigations of the UV metallic lines 
for some Magellanic Cloud B[e] supergiants seen edge-on have revealed terminal 
disk wind velocities of 60--80\,km\,s$^{-1}$ only (Zickgraf et al. 
\cite{Zick96}). They are thus at least a factor of ten
lower than the terminal velocities in the polar winds, which show properties
similar to those of classical B-type star, i.e., line-driven winds (e.g., 
Zickgraf et al. \cite{Zick85}). In combination with the enhanced mass fluxes
found for the equatorial wind, B[e] supergiants show a density contrast of 
100--1000 between the equatorial (i.e. disk) wind and the normal (i.e. polar) 
wind (Zickgraf et al. \cite{Zick89}). Reasonable attempts at explaining these 
equatorial flows are provided by the rotationally induced bi-stability 
mechanism (Lamers \& Pauldrach \cite{LP}; Pelupessy et al. \cite{Pelupessy}; 
Cur\'{e} \cite{Cure04}; Cur\'{e} et al. \cite{Cure05}). 
Even though the formation mechanism is not fully understood and solved yet
(see, e.g., Ignace \& Gayley \cite{Rico}; Kraus \& Miroshnichenko \cite{KM}), 
the disk-forming winds of the B[e] supergiants must be 
dense and cool enough to allow for efficient molecule and dust formation.

The most stable molecule is CO, and during their IR spectroscopic studies of 
early-type emission line stars, McGregor et al. (\cite{McGregor}) detected CO 
band head emission from several targets that later grouped in the 
unclassified B[e] stars or as B[e] supergiant candidates (see 
Table\,\ref{tab_sg}). 

What remains is the question of where in the disk, i.e., at what distance
from the B[e] supergiant star, the CO band emission originates. Recently, Kraus
\& Borges Fernandes (\cite{KrausBorges}) and Kraus et al. (\cite{Krausetal06,
KBA}) have shown that the disks around B[e] supergiants must be predominantly
neutral in hydrogen close to the stellar surface in order to reproduce the
observed strong [O{\sc i}] emission lines. The temperature in the [O{\sc i}]
line forming regions ranges from $\sim 8000$\,K to $\sim 6000$\,K, while the
dust must be located further away from the central star, where the disk
temperatures dropped below the dust evaporation temperature of 1500\,K.
CO has a dissociation temperature of about 5000\,K. Theoretical CO band 
calculations have shown that the formation
of pronounced band head structures requires a minimum CO temperature of about
2000\,K (see, e.g., Kraus \cite{Diplom} and Sect.\,\ref{12CO} below). The CO
bands, therefore, probe the disk area between the [O{\sc i}] and the dust
emission regions. In terms of distance, this temperature argument places the
CO formation region beyond the [O{\sc i}] saturation, which usually happens
at distances of around 100--300\,$R_{*}$ (see Kraus et al. \cite{Krausetal06,
KBA}).

Since the disks around B[e] supergiants are formed from wind 
material, we can expect the disk 
material to mirror the surface composition of the star at the time of ejection. 
Depending on the evolutionary phase of the star at the time of matter ejection, 
the disk material already should be enriched in chemically processed material 
that has been brought to the stellar surface via rotation and mixing processes. 
Searching for such enhanced abundances within the disk material of B[e] stars 
might thus provide us with the information necessary to discriminate 
between a pre- and post-main sequence evolutionary phase of the underlying star.

One of the elements produced during the evolution of massive
stars is the carbon isotope $^{13}$C. Its ratio with the main isotope $^{12}$C
changes drastically during stellar evolution, making $^{13}$C an ideal tracer
for post-main sequence evolutionary phases. We already know of the
existence of $^{12}$CO band emission from the disks of B[e] supergiants, 
but none of the targets has been reported to show $^{13}$CO emission so far. 
The reason for this is certainly the fact that either the spectral resolution
used was generally too low ($R \la 500$, see Sect.\,\ref{basics}), or, when
using high-resolution spectrographs, the spectral region covered by the
observations where restricted to the region around the first $^{12}$CO bandhead
only, which does not include the range of $^{13}$CO emission. This paper, 
therefore, investigates the possible appearance of $^{13}$CO band
head emission during the evolution of OB stars, based on theoretical model 
calculations of the first-overtone bands of $^{12}$CO and $^{13}$CO in 
the $K$ band spectra.

The paper is structured as follows: 
In Sect.\,\ref{evol} we study the variation of the $^{12}$C/$^{13}$C surface
abundance ratio during the evolution of predominantly massive stars. Next, we
calculate simple synthetic CO first-overtone band spectra in 
Sect.\,\ref{basics} of the CO molecule\footnote{Note that by writing CO without 
indication of the isotope, we always refer to the main isotope, i.e., the 
$^{12}$CO molecule.} and discuss the appearance of the band heads of the
$^{13}$CO molecule as a function of $^{12}$C/$^{13}$C ratio as well as of
spectral resolution. In Sect.\,\ref{discussion} we discuss the reliability
of our calculations and applicability to the B[e] supergiants, before
we summarise our results in Sect.\,\ref{conclusions}.

\section{Variation of the $^{13}$C surface abundance during stellar evolution of massive stars}\label{evol}

During the evolution of massive stars, heavy elements produced 
during the different burning processes are brought to the stellar surface via
rotation and mixing processes (see, e.g., the reviews by Pinsonneault 
\cite{Pinsonneault} and Maeder \& Meynet \cite{MaMe}). The surface abundances
of individual elements will thus change with time. Due to the (high) mass loss
of massive stars via line-driven winds, these changes in surface abundances 
will thus translate into abundance changes in the wind, which then clearly 
shows the presence of chemically enriched, processed material. Studies of
the chemical composition of the wind (or more generally: of the circumstellar) 
material of post-main sequence evolutionary phases of massive stars 
are, therefore, ideal to determine the evolutionary phase of the central star.

\begin{figure*}[t!]
\resizebox{\hsize}{!}{\includegraphics{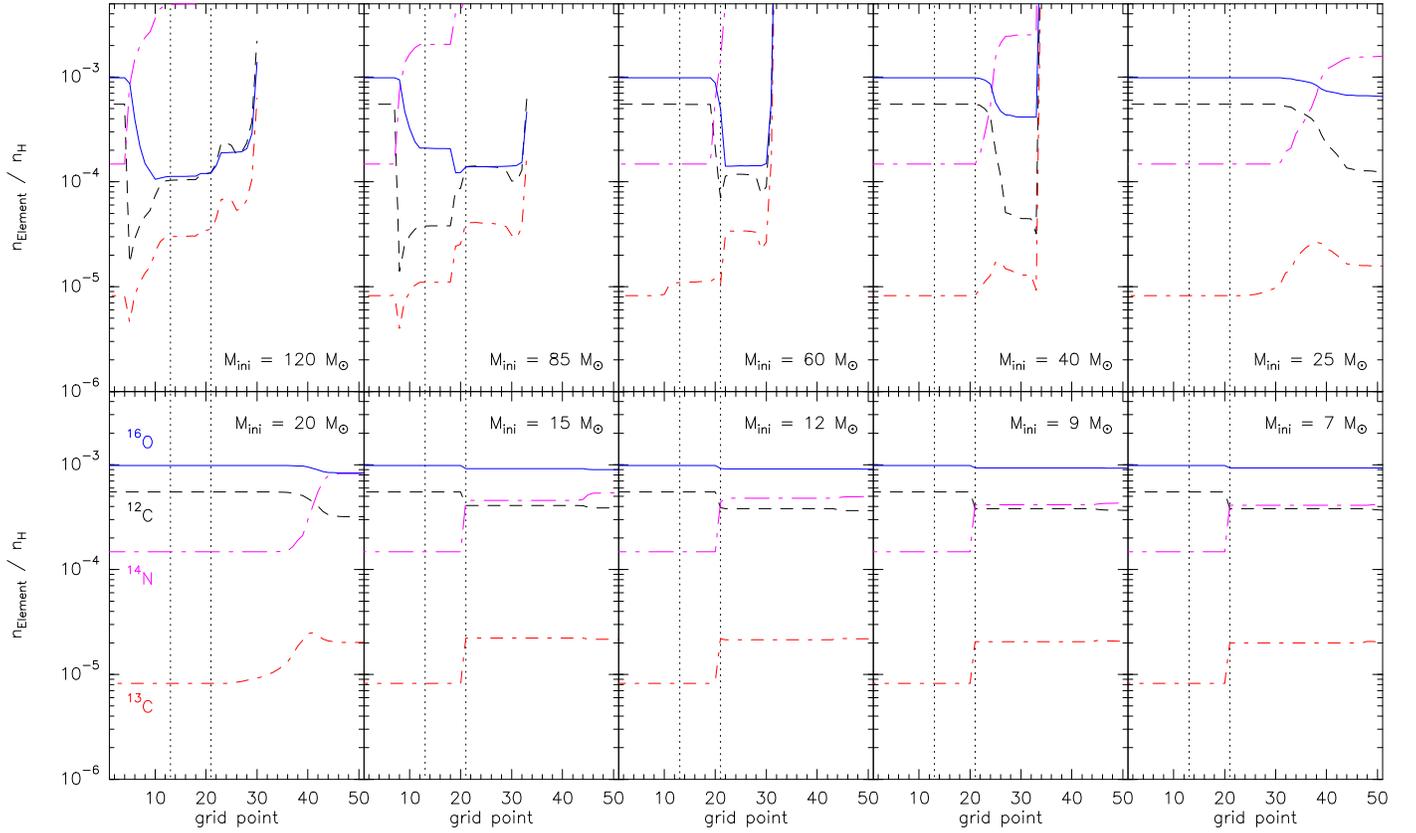}}
\caption{Changes in $^{12}$C, $^{13}$C, $^{14}$N, and $^{16}$O surface 
abundances with respect to the hydrogen surface abundance along the 
evolutionary tracks of non-rotating stars, as indicated by the grid points. The 
ratios were calculated from the data of Schaller et al. (\cite{Schaller}) for 
solar metallicity. The approximate end of the core H burning and the onset of 
the core He burning phases are indicated by the two vertical dotted lines in 
each panel.} 
\label{ab}
\end{figure*}

\begin{figure*}[t!]
\resizebox{\hsize}{!}{\includegraphics{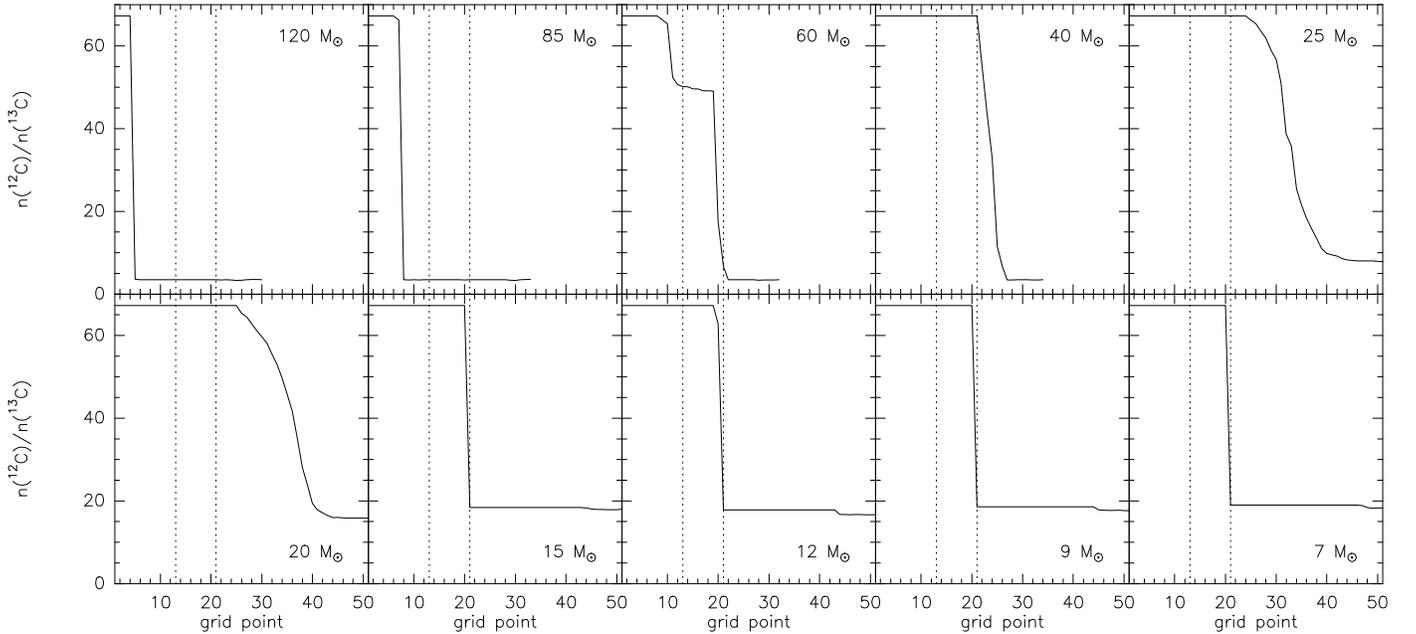}}
\caption{Change of the $^{12}$C/$^{13}$C surface abundance ratio during stellar
evolution of non-rotating massive stars at solar metallicity. The ratios are
calculated from the tables of Schaller et al. (\cite{Schaller}). The dotted
lines have the same meaning as in Fig.\,\ref{ab}.}
\label{ratio}
\end{figure*}

To study the changes in surface abundances (and hence of the wind abundances),
we use the grid of Schaller et al. (\cite{Schaller}) for non-rotating stars at 
solar metallicity. In Fig.\,\ref{ab} we plot the changes in surface abundances 
along the evolutionary tracks (indicated by the grid numbers) of $^{12}$C,
$^{13}$C, $^{14}$N, and $^{16}$O with respect to the surface hydrogen abundance 
at the same grid point. The ratios have been calculated from the mass fractions 
provided in the data files of Schaller et al. (\cite{Schaller}). The grid 
points that indicate the approximate end of the core H burning and the onset of 
the core He burning phases are indicated by the two vertical dotted lines in 
each panel. The abundance curves for stars with initial masses $M_{\rm ini} > 
40$\,M$_{\odot}$ stop around grid point $\sim 30$ (i.e. before the end of core 
He burning) since these stars have lost their complete outer hydrogen shell via 
strong mass loss.

The change in chemical surface composition during the main-sequence 
lifetime is quite obvious for the most massive stars, while the surface 
abundances of the less massive stars remain at their initial values until the 
onset of core He burning or even longer. The elements with clearly increasing 
abundances during the evolution are $^{14}$N and $^{13}$C, while at the same 
time the abundances of $^{16}$O and $^{12}$C slightly decrease. The 
nitrogen enrichment of the surface (and hence the wind) material is usually 
used as an indicator to trace evolved evolutionary phases of massive stars, as 
for the classical OB supergiants (see, e.g., Crowther et al. \cite{Crowther}; 
Hillier et al. \cite{Hillier}; Evans et al. \cite{Evans}; Searle et al. 
\cite{Searle}) or the Luminous Blue Variables (e.g., Lamers et al. 
\cite{Lamers}).

The direct study and comparison of the surface abundances of the B[e] stars 
would be an ideal tool for the determination of their evolutionary phases. 
Unfortunately, due to their usually high density circumstellar material (caused
by high mass loss in the case of B[e] supergiants), their spectra are often 
crowded with pure emission lines, and the photosphere of most targets is not 
visible at all. A direct way to measure their surface abundances is thus not 
possible. The only way to draw conclusions about their surface composition is 
to study the chemical abundances from their emission lines formed within their 
close-by circumstellar material. However, to determine the CNO abundances of 
B[e] stars from their emission lines, it is necessary to have a detailed 
description of the exact geometry, the inclination angle of the system, and the 
latitude dependent ionization structure. This information is usually not 
available and can hardly be derived from the modeling of the emission lines 
alone. We therefore need another, more reliable indicator for the 
evolutionary state of an individual object.

One important indicator of post-main 
sequence evolutionary phases is the surface abundance of the isotope $^{13}$C, 
or more precisely, the change in the isotopic ratio $^{12}$C/$^{13}$C. The 
strong increase in $^{13}$C in combination with the (slight) decrease in 
$^{12}$C surface abundances (see Fig.\,\ref{ab}) results in a rather steep 
decrease of the $^{12}$C/$^{13}$C ratio from its initial interstellar value of
$\sim 70$ to values $< 20$ for stars with initial masses $\la 40$\,M$_{\odot}$, and for stars with initial masses $\ga 60$\,M$_{\odot}$ even to 
$^{12}$C/$^{13}$C $\la 5$. The variation in the $^{12}$C/$^{13}$C surface 
abundance ratio during the evolution of massive stars at different metallicity 
is depicted in Fig.\,\ref{ratio}.

With an abundance ratio lower than about 20, the $^{13}$C isotope should be
easily detectable. However, our intension is not to search for individual 
emission lines of the $^{13}$C atom,
but rather to concentrate on the molecular emission of CO and its $^{13}$CO 
isotope.
Besides, our investigations will be restricted to the $^{12}$CO and $^{13}$CO, 
only, since the abundances in $^{17}$O and $^{18}$O compared to the main 
isotope $^{16}$O are generally extremely low, even during the late evolutionary 
phases of massive stars. Contributions from the C$^{17}$O and C$^{18}$O 
isotopes to the total CO band structures are thus negligible.

\section{CO band formation}\label{basics}

CO first-overtone band emission is a well known feature arising especially from 
the rotating accretion disks around young stellar objects (YSO), and the 
physics of CO band formation in such rotating pre-main sequence disks has 
already been discussed widely in the literature (see, e.g., Scoville et al. 
\cite{Scoville}; Carr \cite{Carr89}; Najita et al. \cite{Najita}; Kraus 
\cite{PhD}; Kraus et al. \cite{Kraus00}). The main focus in YSO research is 
therefore especially the structure of the $2\longrightarrow 0$ band head, which 
contains the complete velocity information, and from which the rotation 
velocity, disk inclination, as well as some information concerning the density 
and temperature within the CO forming disk region can be derived (e.g., Carr et 
al. \cite{Carr93}; Chandler et al. \cite{Chandler}; Carr \cite{Carr95}; Najita 
et al. \cite{Najita}; Kraus \cite{PhD}; Kraus et al. \cite{Kraus00}; Bik \& Thi
\cite{BikThi}; Berthoud et al. \cite{Berthoud}).

Such high-resolution $2\longrightarrow 0$ band head observations also would 
be very important for B[e] stars in studying the kinematics of the CO gas, 
especially for the discrimination between whether the disks around B[e] 
supergiants are Keplerian rotating (i.e., stable, long-lived disks), or 
outflowing disk-forming winds (see, e.g., Porter \cite{John}; Kraus et al. 
\cite{KBA}). For our purpose related to the B[e] supergiants' evolutionary 
phase, however, we need to study a much larger (in wavelength) portion of the 
complete CO first-overtone band structures, because we want to find out (i) 
where in the $K$ band spectra of B[e] supergiants we can expect the $^{13}$CO 
band heads to show up, (ii) during which evolutionary phases observable 
$^{13}$CO band head emission might be generated, and (iii) what the most 
suitable spectral resolution is to easily detect these $^{13}$CO band heads.

\subsection{Synthetic band spectra of $^{12}$CO}\label{12CO}

For the calculation of the CO band spectra, we follow the description of Kraus 
et al. (\cite{Kraus00}). Due to the lack of knowledge of the proper shape
and structure of the B[e] supergiant stars' disks, we keep the calculations
simple, assuming the CO gas to be isothermal. We further assume that the CO 
gas is in local thermodynamic equilibrium. Then, the emission at frequency 
$\nu$ is simply given by 
\begin{equation}
I_{\nu} = B_{\nu}(T_{\rm CO}) (1 - e^{-\tau_{\nu}})
\label{int}
\end{equation} 
with the optical depth
\begin{equation}
\tau_{\nu} = \kappa_{\nu, ^{12}\rm CO} N_{^{12}\rm CO}\,.
\label{tau}
\end{equation}
Here, $\kappa_{\nu, ^{12}\rm CO}$ is the absorption coefficient per $^{12}$CO 
molecule, and $N_{^{12}\rm CO}$ is the column density of the $^{12}$CO gas for 
which we adopt a value of $10^{17}$\,cm$^{-2}$ in order to keep the emission 
optically thin at all wavelengths over the complete band spectrum. 
We take into account a total of 10 vibration bands,
and within each vibration band a total of 140 rotation levels. The so-called
Dunham coefficients to calculate the energy levels are taken from Farrenq et
al. (\cite{Farrenq}), and the Einstein coefficients for spontaneous emission
from Chandra et al. (\cite{Chandra}).

To calculate $\kappa_{\nu, ^{12}\rm CO}$, we need to specify the profile 
function of the individual vibration-rotational lines for which we use, for
simplicity, a Gaussian profile. The width of the Gaussian 
we describe by a fixed (internal) CO velocity of 30\,km\,s$^{-1}$. The choice
of this velocity is not crucial, unless the internal velocity exceeds
the one given by the spectral resolution. For some first test calculations, 
we smear the CO band spectra to an arbitrary spectral resolution of 
$R = 1500$, corresponding to a velocity resolution of about 200\,km\,s$^{-1}$.

\begin{figure}[t!]
\resizebox{\hsize}{!}{\includegraphics{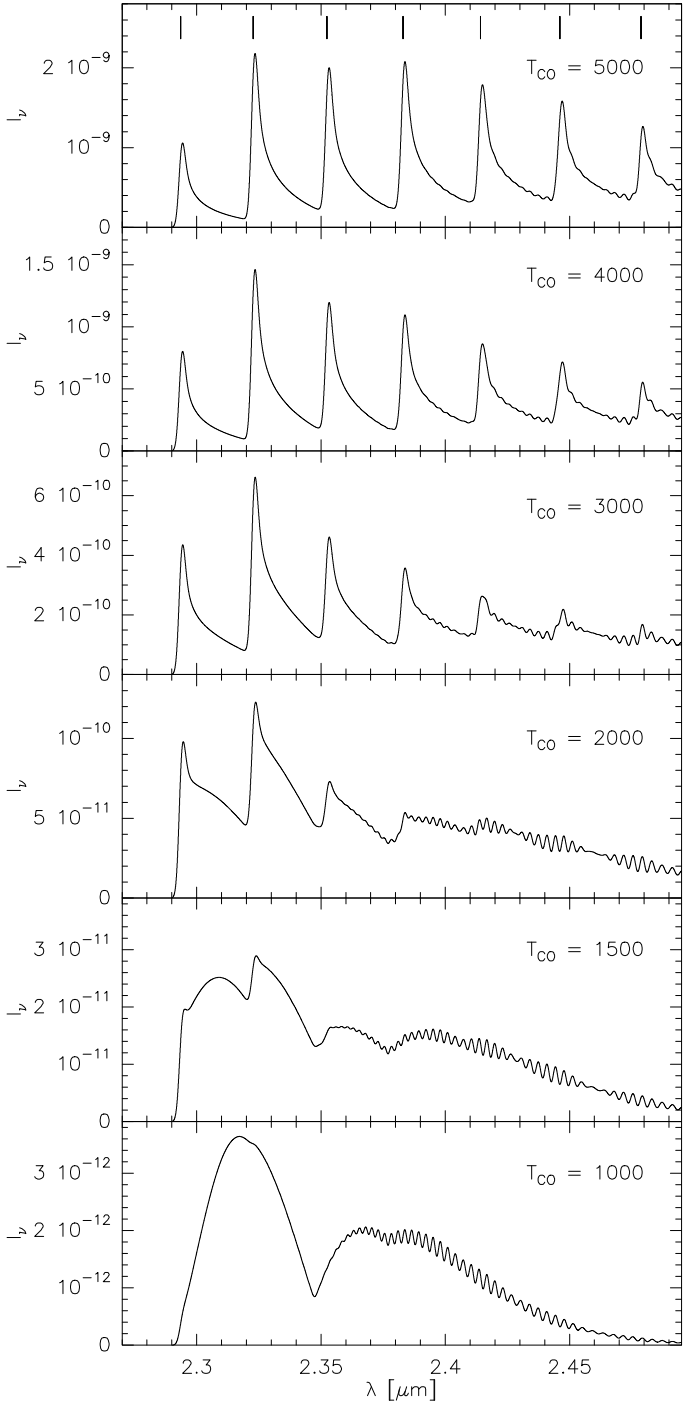}}
\caption{Influence of the temperature on the synthetic $^{12}$CO band emission
spectra. Clear band head structures form for $T_{\rm CO} \ge 2000$\,K.}
\label{temp}
\end{figure}

To study the variation in the individual band head peak intensities, we  
calculate the structure of the first-overtone bands for different CO 
temperatures, ranging from 1000\,K to 5000\,K. The results are shown in 
Fig.\,\ref{temp} with the positions of the band edges\footnote{For 
clarification, we use the terminus {\it band head} when talking about the 
intensity peaks, while the transition of lowest wavelength in each band we call 
the {\it band edge}.} indicated by the ticks in the top panel. The formation of 
clear band heads starts for temperatures $T_{\rm CO}\ga 2000$\,K, while for CO 
temperatures below 2000\,K no individual and clearly identifiable band heads 
emerge. For $T_{\rm CO}\simeq 1000$\,K only the levels within the first 
vibrational band are excited,
giving rise to the broad, double emission feature. The higher the temperature, 
the more vibrational bands can be excited and the more individual band heads 
appear. A further effect of incresing CO temperature is the strong increase in 
band head intensities, with the $3\longrightarrow 1$ as the clearly dominating 
one, while the $2\longrightarrow 0$ band head becomes depressed with increasing 
temperature due to depopulation into the higher vibrational levels. Due to this 
strong intensity increase with CO temperature, most of the observable band head 
emission originates from the hottest CO gas, i.e., from the disk regions 
closest to the star. 

\begin{figure*}[t!]
\resizebox{\hsize}{!}{\includegraphics{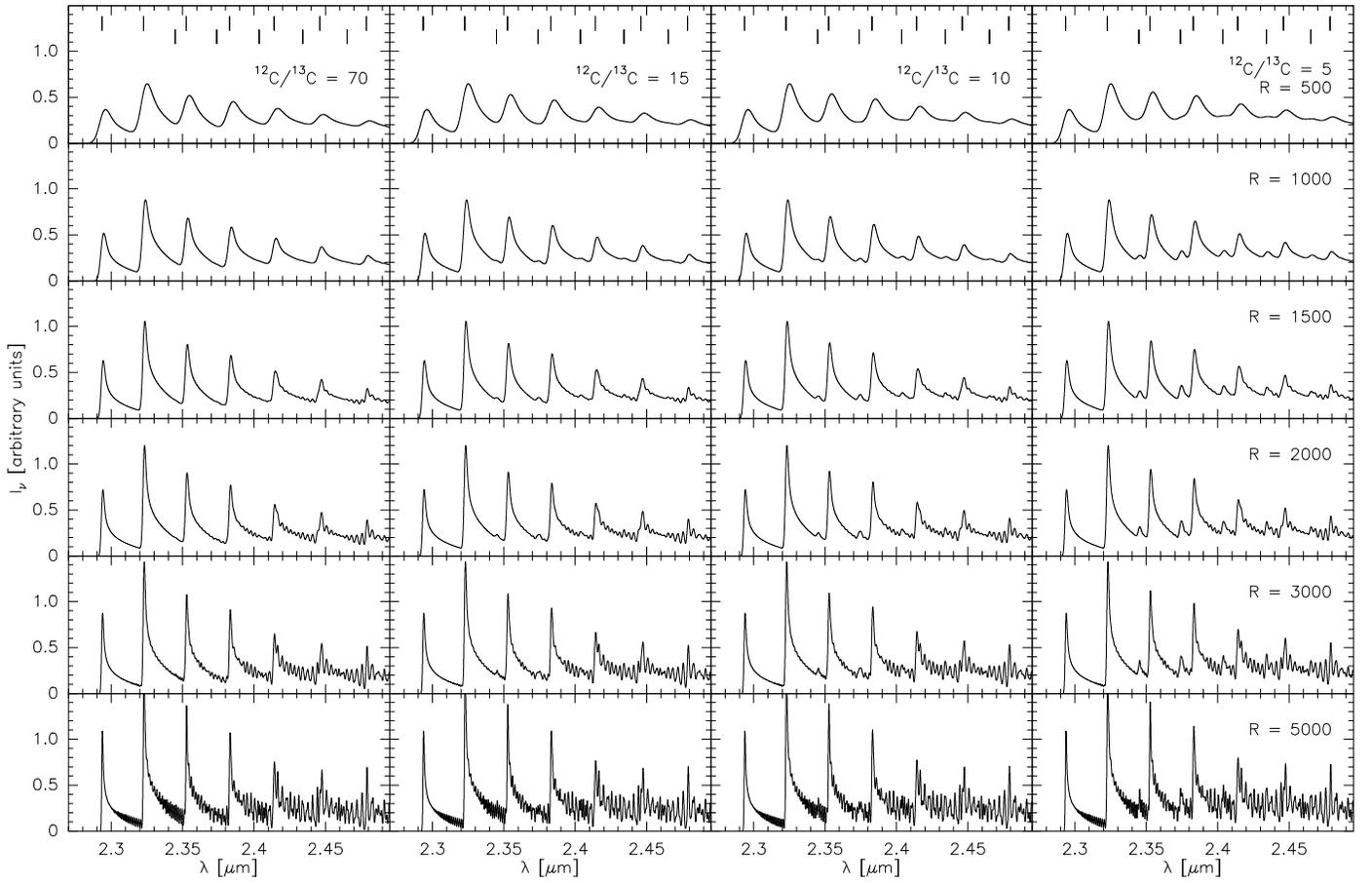}}
\caption{Synthetic, optically thin $^{12}$CO plus $^{13}$CO first-overtone band 
spectra computed for $T_{\rm CO} = 3500$\,K. The individual panels are for 
increasing spectral resolutions (from top to bottom) and for decreasing
$^{12}$C/$^{13}$C ratios (from left to right).}
\label{combine}
\end{figure*}

\subsection{The appearance of the $^{13}$CO bands}\label{13CO}

The slightly higher mass of the $^{13}$C isotope has two consequences 
for the energy levels within the diatomic molecule: (i) A higher mass
results in a higher moment of inertia, reducing the energy of the individual
rotational levels within each vibrational band.
(ii) Since the frequencies of the vibrations depend on the reduced mass only, 
the vibrational energy levels within the more massive $^{13}$CO isotope will 
all be reduced compared to those within the $^{12}$CO molecule. Consequently, 
the net effect of the higher mass of the isotope is a shift of the coupled 
vibration-rotational transitions, and hence of the complete band spectrum, to 
longer wavelengths. Since the wavelength shift (i.e., the reduction in energy) 
is especially strong for the vibrational transitions, a clear wavelength 
separation appears between the bands of $^{12}$CO and $^{13}$CO. 
The lowest band edge of the first-overtone bands arises at 
2.3448\,$\mu$m ($2\longrightarrow 0$), i.e., at a wavelength between the second 
and third band edge of the $^{12}$CO bands. This fact is of major importance 
for the identification of the $^{13}$CO band heads in the total spectrum.

Since the main effect of the different masses is a shift of the individual
vibration-rotational lines to higher wavelengths, the bands of the $^{13}$CO 
molecule will, in general, have a very similar shape to those of $^{12}$CO. To 
compute the total band spectrum consisting of the $^{12}$CO and $^{13}$CO 
transitions, we again calculate the intensity, $I_{\nu}$, at each wavelength 
with Eq.\,(\ref{int}), but for the optical depth, $\tau_{\nu}$, we now need to 
include the contribution of the $^{13}$CO molecule into Eq.\,(\ref{tau}), 
giving 
\begin{equation}
\tau_{\nu} = \kappa_{\nu, ^{12}\rm CO} N_{^{12}\rm CO}+\kappa_{\nu, ^{13}\rm 
CO} N_{^{13}\rm CO}\, ,
\end{equation} 
where $N_{^{13}\rm CO}$ and $\kappa_{\nu, ^{13}\rm CO}$ are the column density
and the absorption coefficient per $^{13}$CO molecule, respectively.

It is convenient to parametrize the column density of the isotope, $N_{^{13}\rm 
CO}$, in terms of $N_{^{12}\rm CO}$ and the carbon isotope abundance ratio, 
$n_{^{12}\rm C}/n_{^{13}\rm C}$, for which we assume that it equally translates 
into the CO isotope ratio, so that we can write  
\begin{equation}
N_{^{13}\rm CO} = \frac{N_{^{13}\rm CO}}{N_{^{12}\rm CO}} N_{^{12}\rm CO}
= \frac{n_{^{13}\rm CO}}{n_{^{12}\rm CO}} N_{^{12}\rm CO}
= \frac{N_{^{12}\rm CO}}{n_{^{12}\rm C}/n_{^{13}\rm C}}\,,
\end{equation}
with the carbon isotope abundance ratio, $n_{^{12}\rm C}/n_{^{13}\rm C}$, as
the only free parameter. From this relation it is already obvious that
the emergence of the $^{13}$CO bands strongly depends on the carbon isotopic
ratio, i.e., the lower this ratio, the higher the probability of $^{13}$CO band 
appearance.

In our test calculations we use the following values for the $n_{^{12}\rm C}/ 
n_{^{13}\rm C}$ abundance ratio: 70, 15, 10, and 5. While the first value 
approximately represents the interstellar (i.e., initial) ratio, 
the latter appear during the more interesting, later evolutionary stages 
of stars with different initial masses (see Fig.\,\ref{ratio}). The emerging
total spectra for these four ratios and for different spectral resolutions
are displayed in Fig.\,\ref{combine}, with the wavelengths of both, the 
$^{12}$CO and $^{13}$CO band edges, indicated in the top panels.

For an interstellar carbon isotopic ratio of 70, the $^{13}$CO bands remain 
undetectable. The first indications for $^{13}$CO band heads peaking out 
of the total CO spectrum are for an isotopic ratio of 15, while for
even lower ratios, the individual $^{13}$CO band heads are clearly present.
However, the appearance of these band heads strongly depends on the chosen
spectral resolution. To show the effect of spectral resolution, we calculated
synthetic CO bands for increasing resolution, from $R = 500$, which corresponds
approximately to the value used by McGregor et al. (\cite{McGregor}) as was 
discussed in Sect.\,\ref{intro}, to $R = 5000$, with the intrinsic CO
velocity fixed at 30\,km\,s$^{-1}$. While for the lowest resolution the 
$^{13}$CO band heads are not visible at all, their smooth structure breaks down
at the chosen highest resolution, since at high resolution we start to resolve 
individual vibration-rotational lines. The most suitable range in 
resolution that follows from these calculations is thus between $R = 1000$ and 
$R = 3000$.

The terminal velocity of the equatorial winds in B[e] supergiants was found to 
be $\la 80$\,km\,s$^{-1}$ (Zickgraf et al. \cite{Zick96}), so that for 
edge-on systems a resolution as high as $R = 5000$ would also still
deliver smooth $^{13}$CO band heads. However, since not much is known about the
possible inclinations of the galactic B[e] supergiant candidates, a lower 
resolution ($R \la 3000$) is preferable for the detection of smooth 
$^{13}$CO band heads.

\subsection{Influence of optical depth effects}\label{optdick}
                                                                                
Besides the CO temperature, the column density of the emitting region is
a further important parameter that triggers the intensity (and thus
the observability) of the CO bands. In this section, we therefore briefly
discuss the influence of the CO column density, and hence of the optical
depth, on the global appearance of the CO bands and, especially, on the 
appearance of the $^{13}$CO band heads.

\begin{figure}[t!]
\resizebox{\hsize}{!}{\includegraphics{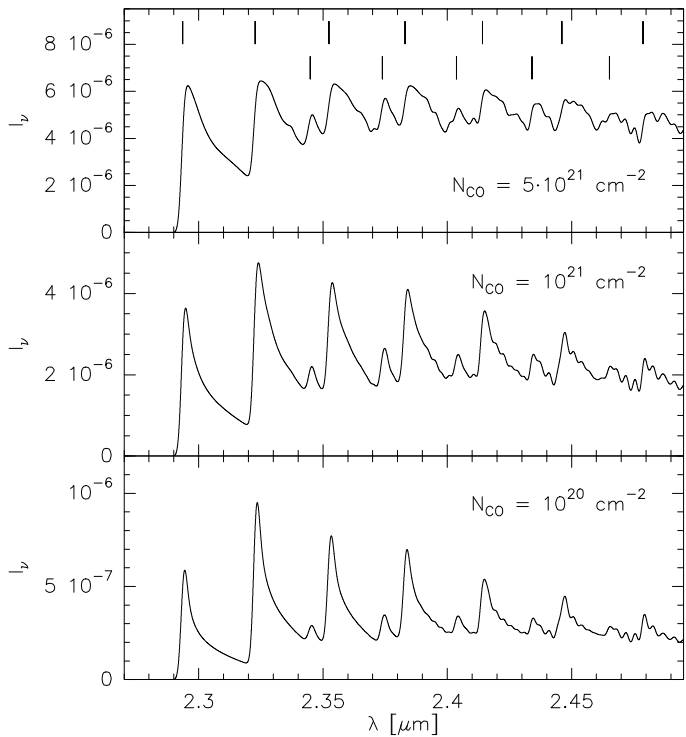}}
\caption{CO Band structure for increasing CO column density (from bottom to
top), i.e., increasing optical thickness.}
\label{thick}
\end{figure}

We calculate the CO bands for a fixed temperature of $T_{\rm CO} = 3500$\,K
and carbon isotopic ratio of $^{12}$C/$^{13}{\rm C} = 5$ and for increasing 
$^{12}$CO column densities. The results for three different values of the 
column density are shown in Fig.\,\ref{thick}. Contrary to the mid and top 
panel, which show clear indications for partially optically thick emission, the 
chosen column density in the bottom panel is low enough to guarantee that the 
CO bands are still optically thin throughout the spectrum.

The influence of the increasing optical depth becomes obvious in the strong
increase of the CO "quasi-continuum", starting especially at long wavelengths, 
which leads to a reduction of the ratio between the peak and the
inter-peak (i.e., the quasi-continuum) intensity. An additional optical depth 
effect is the broadening of the band heads on their long-wavelength "wing". 
Nevertheless, the $^{13}$CO band heads remain detectable, even for significant
optical depth values within the $^{12}$CO bands.

\section{Discussion}\label{discussion}

The rising of the $^{13}$CO band heads with decreasing carbon isotopic ratio is 
nicely displayed in Fig.\,\ref{combine}. Since the decrease in isotopic ratio 
is a clear indicator of stellar evolution (see Fig.\,\ref{ratio} and 
Sect.\,\ref{evol}), we can claim that, if $^{13}$CO band head 
emission is identified within the $K$ band spectra of a massive star, it 
hints at an evolved nature of the object. Whether and how this 
can be used to solve the problem of the evolutionary phase of the B[e] 
supergiants, or B[e] stars in general, is discussed in the following.

\subsection{Comparison with non-rotating stellar evolution models}\label{predict}

\begin{figure*}[t!]
\resizebox{\hsize}{!}{\includegraphics{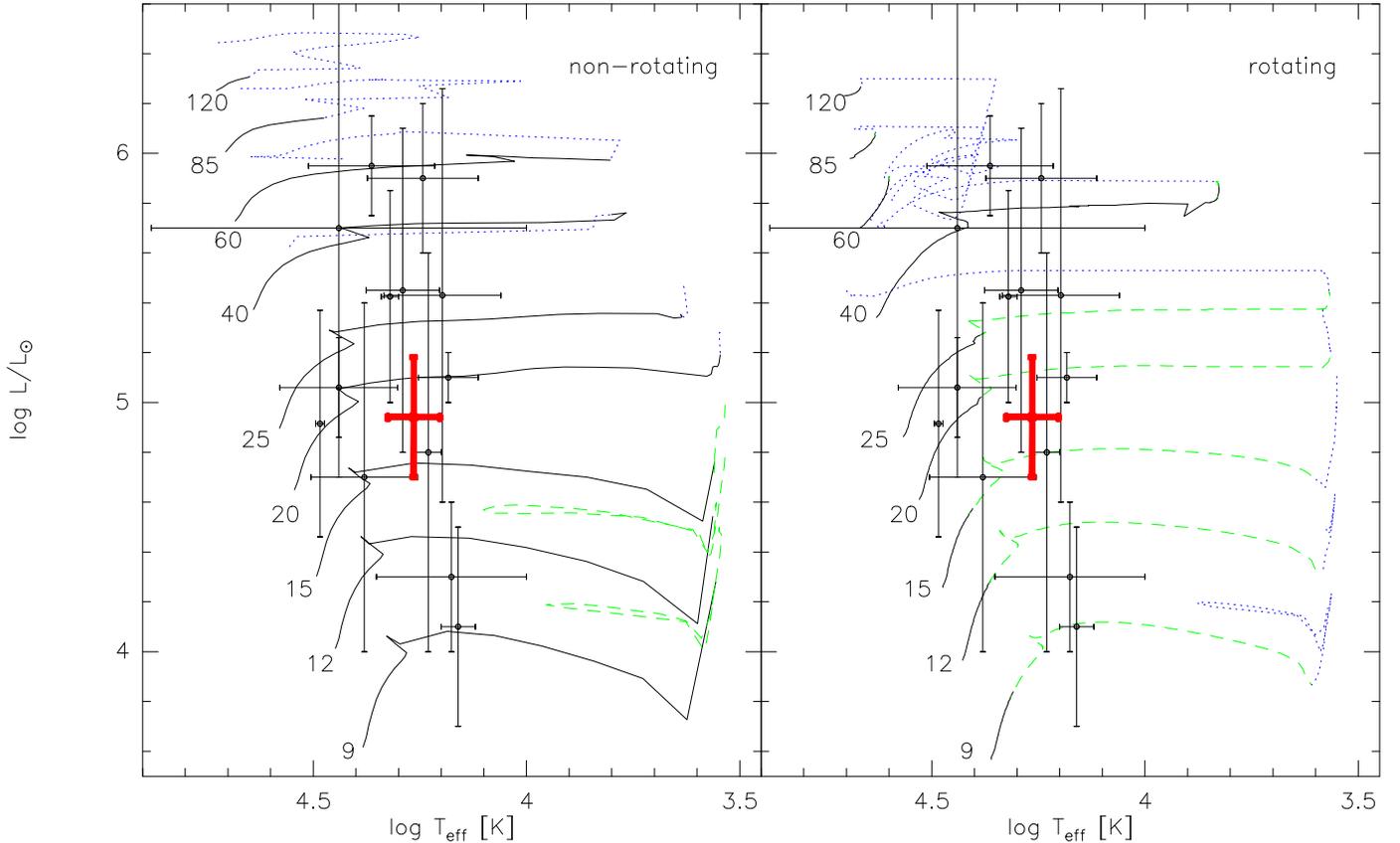}}
\caption{Comparison of the positions of the galactic B[e] supergiant 
candidates from Table\,\ref{tab_sg} with evolutionary tracks of non-rotating 
(left panel, from Schaller et al. \cite{Schaller}) and rotating (right panel, 
from Meynet \& Maeder \cite{MeynetMaeder}) stellar evolutionary models at 
solar metallicity. The dashed and dotted parts of the tracks indicate those 
evolutionary phases with $20 \ge ^{12}$C/$^{13}{\rm C} \ge 15$ and
$^{12}$C/$^{13}{\rm C} \le 15$, respectively. The position of GG\,Car is 
emphasized.}
\label{hrd}
\end{figure*}

For a sense of whether the B[e] supergiant candidates in Table\,\ref{tab_sg} 
will show $^{13}$CO bands in their $K$ band spectra at all, we need to specify 
those evolutionary phases with suitably low isotopic ratio and compare them 
with the locations of the B[e] stars within the HR diagram. Such a comparison 
is performed in the left panel of Fig.\,\ref{hrd}, where we plot the stellar 
evolutionary tracks from Schaller et al. (\cite{Schaller}) for non-rotating
stars at solar metallicity. Those evolutionary phases during which the 
$^{12}$C/$^{13}$C ratio drops below $\la 20$ and further below $\la 15$ are 
indicated. Also included are the positions (with error bars)
of the galactic B[e] supergiant candidate sample.

Obviously, only the evolutionary tracks for stars with highest initial mass 
(i.e., $M_{\rm in} \ga 40\,M_{\odot}$) show an overlap of their evolutionary 
phases of low carbon isotopic ratio with the positions of the most luminous
B[e] supergiant candidates. For these targets, the detection of $^{13}$CO bands 
would immediately classify these as B[e] supergiants, and for stars in the
mass range $40\,M_{\odot} \la M_{\rm in} \la 60\,M_{\odot}$, even constrain the 
B[e] supergiant phase to late evolutionary phases, i.e., within the second
blue supergiant phase during the evolution.

Stellar models with $M_{\rm in} < 15\,M_{\odot}$ reach a carbon isotopic ratio 
of about 15--20 during their evolution along the blue loop. Even though there 
exist some B[e] stars within this mass range, the $^{12}$C/$^{13}$C ratio for 
these stars might not be low enough for a definite $^{13}$CO band detection in 
low resolution. For these stars, a high-resolution high signal-to-noise 
spectrum covering the region around the $2\rightarrow 0$ $^{13}$CO band head 
might be more suitable to search for (and resolve) individual 
vibration-rotation lines.

For the stellar models of intermediate mass stars ($15\,M_{\odot}
\la M_{\rm in} \la 40\,M_{\odot}$), the $^{12}$C/$^{13}$C ratio changes 
only at the end of their evolution, i.e. within the red part of the HR diagram.
For this mass range thus there exists no overlap of the low carbon isotopic 
ratio regime with the locations of the B[e] supergiant candidates.

\subsection{The influence of stellar rotation}
                                                                                
It is generally assumed that the disk formation around B[e] supergiants is 
linked to rapid rotation of the central stars. In fact, for 4 Magellanic Cloud 
B[e] supergiants, projected stellar rotation values (i.e., $\varv\sin i$) could 
be determined. These stars indeed seem to rotate at a significant fraction of 
their critical velocity (Gummersbach et al. \cite{Gummersbach}; Zickgraf 
\cite{Zickgraf00, Zickgraf06}; Kraus et al. \cite{Kraus08}). Whether all B[e] 
supergiants are rapid rotators, is, however, not known and difficult to 
determine, since for the rest of the targets no photospheric absorption lines 
can be detected due to their high-density winds. 

To date, nothing is known about possible rotation velocities of the galactic 
B[e] supergiant candidates. Nevertheless, we investigated the influence of 
stellar rotation on the variation of the surface $^{12}$C/$^{13}$C isotope 
ratio. For this, we used the tables of Meynet \& Maeder (\cite{MeynetMaeder}) 
for rotating stars at solar metallicity. The evolutionary tracks are shown in 
the right panel of Fig.\,\ref{hrd}, and the evolutionary phases with $20 \ge 
^{12}$C/$^{13}{\rm C} \ge 15$ and $^{12}$C/$^{13}{\rm C} \le 15$ are again 
indicated. The general trend of the stellar rotation is to enhance the stellar 
surface abundance with $^{13}$C much earlier during the stellar evolution. And 
interestingly, for the rotating models the $^{12}$C/$^{13}$C isotope ratio
drops below a value of 20 already during the main-sequence evolution of even
the stars with lowest (i.e., here 9\,$M_{\odot}$) initial mass, and even stars
with an initial mass of 25\,$M_{\odot}$ experience a decrease in isotope ratio 
to $^{12}$C/$^{13}{\rm C} \le 15$ during their late evolutionary phase. In the 
presence of stellar rotation, the positions of all our B[e] supergiant 
candidates thus fall into regions with decreased $^{12}$C/$^{13}$C isotope 
ratio. The search for $^{13}$CO band emission, therefore, seems to be 
a promising way to distinguish between pre- and post-main sequence 
evolution of B[e] stars.

\subsection{The B[e] supergiant candidate GG\,Car}

In Fig.\,\ref{hrd} we emphasized the position of the B[e] supergiant
candidate \object{GG\,Car} based on the stellar parameter determinations of
earlier investigations by McGregor et al. (\cite{McGregor}) and Lopes et al.
(\cite{Lopes}). According to these studies, it seems that GG\,Car has evolved
from a progenitor star of $M_{\rm in} \simeq 15\ldots 22\,M_{\odot}$.

Recently, Domiciano de Souza et al. (in preparation) performed VLTI/AMBER 
observations of GG\,Car and extracted its $K$ band spectrum. Interestingly, the 
AMBER spectrum of GG\,Car, which has a spectral resolution of $R\sim 1500$, 
seems to display clearly detectable $^{13}$CO band emission (Chesneau \& 
Domiciano de Souza, private com.), classifying GG\,Car as a B[e] supergiant 
star. Based on the peak/inter-peak intensity ratio, the CO emission seems to 
be rather optically thin, and
the detected strength of the $^{13}$CO band heads speaks in favour of a 
$^{12}$C/$^{13}$C isotope ratio of the order of $\la 10$. Such a low ratio is, 
however, not expected within the suspected evolutionary phase of GG\,Car, 
neither from the non-rotating, nor from the rotating stellar evolution models. 
This discrepency might indicate that either the luminosity of GG\,Car (and 
hence its initial mass) has been considerably underestimated, or its progenitor 
star must perhaps have been rotating much faster than 300\,km\,s$^{-1}$, 
resulting in a much stronger $^{13}$C surface abundance enhancement than the 
predictions from the rotating models of Meynet \& Maeder (\cite{MeynetMaeder}). 
In any case, the detection of the unexpectedly strong $^{13}$CO band heads in 
GG\,Car needs to be studied in more detail, including a careful modeling of the 
$^{12}$CO and $^{13}$CO band emission (Domiciano de Souza et al., in 
preparation). In addition, more interferometric studies to spatially resolve 
the CO bands, as has recently been shown by Tatulli et al. (\cite{Tatulli}), 
are definitely needed.

\section{Conclusions}\label{conclusions}

For many galactic B[e] stars it is not clear whether they are pre-main
sequence stars (i.e., Herbig stars) or whether they belong to the group
of B[e] supergiants. The main reason for this long-standing problem in
classification is their often only poorly known distance, resulting in
huge errorbars for the stellar luminosities. In addition, these stars usually
have high density circumstellar material obscuring the stellar photosphere,
so that an abundance study of their chemical surface composition is not
possible.

To overcome these problems, we suggest a different way to distinguish
between the pre- and post-main sequence evolutionary phase of the individual
targets based on the increasing $^{13}$C abundance on the stellar surface at
post-main sequence evolutionary phases, and hence an increased $^{13}$C 
abundance in the winds of the supergiant stars. Since typical B[e] supergiants
(such as those in the Magellanic Clouds) show the presence of
CO band emission arising in their high-density, neutral circumstellar disks,
we investigate the conditions under which observable $^{13}$CO band emission
in the $K$ band spectra of B[e] supergiant candidates is generated. Only stars 
for which the $^{12}$C/$^{13}$C isotope ratio on the stellar surface (and hence 
in the wind material) has dropped substantially, from the initial, interstellar 
value of about 70 to less than 20, are found to be suitable to produce 
observable $^{13}$CO band emission. Such a drop in isotope ratio happens
naturally in the post-main sequence evolution of extremely massive stars, for 
which mixing processes and the loss of the outer shells via strong mass loss
result in a strong increase in $^{13}$C surface abundance. In the case of
rotating stellar models, this surface enrichment in the $^{13}$C isotope
is much more efficient so that even stars with moderate initial mass (i.e.,
stars with $M_{\rm in} \ga 9\,M_{\odot}$) already show considerable variations,
i.e., a drop in the $^{12}$C/$^{13}$C isotope ratio. The computation of 
synthetic emission spectra of both the $^{12}$CO and $^{13}$CO bands for 
different resolutions revealed that a spectral resolution of $R\sim 1500\ldots
3000$ is ideal for the detection of the $^{13}$CO band heads.

Interestingly, the B[e] supergiant candidate GG\,Car has recently been found to 
show $^{13}$CO band emission, classifying it definitely as a B[e] supergiant. 
The observed strength of its $^{13}$CO bands is, however, in disagreement with 
the model predictions for its postulated position within the HR diagram, 
meaning that either its luminosity and initial mass have been considerably 
underestimated, or its progenitor star must have been rotating much faster than 
the 300\,km\,s$^{-1}$ used in the rotating stellar evolution models. And 
also for the B[e] supergiant candidate CPD-57$\degr$2874 some tentative 
$^{13}$CO detection has been reported (Chesneau \& Domiciano de Souza, private
com.), indicating that the search for and detection of $^{13}$CO bands is 
indeed an ideal tool to discriminate between the pre- and post-main sequence 
evolutionary phase for the still large number of unclassified B[e] stars.


\begin{appendix}

\section{Galactic B[e] supergiant candidates}

Table\,\ref{tab_sg} contains a list of the currently known galactic B[e] supergiant candidates.
 
\begin{table*}
\caption{Parameters of galactic B[e] supergiant candidates. For each object we
provide the ranges in luminosity and temperature as found in the literature.
We further indicate for each target whether CO band emission has been detected 
in their $K$ band spectra. A question mark means that no CO band emission was 
detected for $R\sim 450$.}
\begin{tabular}{lcclccl}
\hline
\hline
Object & $\log L/L_{\odot}$ & $T_{\rm eff}$ [K] & Ref. & $K$ band & $^{12}$CO  & Ref. \\
       &                    &                   &      & spectra  &   & \\
\hline
CD-42$\degr$11721 & $4.0\pm0.3$   & 13--15\,000   & Borges Fernandes et al. (\cite{Marcelo}) & yes & ? & McGregor et al. (\cite{McGregor})\\
                  & $>4.5$        & 16\,000       & McGregor et al. (\cite{McGregor}) & & & \\[0.3em]
CPD-52$\degr$9243 & 5.9           & 16\,200       & Lopes et al. (\cite{Lopes}) & yes & yes & McGregor et al. (\cite{McGregor}) \\
                  & $5.45\pm0.25$ & 13--22\,000   & Winkler \& Wolf (\cite{WinklerWolf}) & & & \\
                  & 5.4           & 16\,000       & McGregor et al. (\cite{McGregor}) & & & \\[0.3em]
CPD-57$\degr$2874 & $> 4.8$       & 16\,000       & McGregor et al. (\cite{McGregor}) & yes & yes & McGregor et al. (\cite{McGregor}); \\
                  & $5.7\pm0.4$   & 17--23\,000   & Domiciano de Souza &  &  & Domiciano de Souza \\
                  &               &               & et al. (\cite{Domiciano})& &  & et al. (in prep.) \\[0.3em]
GG\,Car           & 4.7           & 16\,000       & McGregor et al. (\cite{McGregor}) & yes & yes & McGregor et al. (\cite{McGregor}); Morris \\
                  & 5.18          & 20\,800       & Lopes et al. (\cite{Lopes}) &  &  & et al. (\cite{Morris}); Domiciano de\\
                  &               &               &  &  &  & Souza et al. (in prep.)\\[0.3em]
Hen 3-298         & $5.1\pm0.3$   & 13--17\,500   & Miroshnichenko et al. (\cite{Mirosh05}) & yes & yes & Miroshnichenko et al. (\cite{Mirosh05})\\[0.3em]
Hen 3-303         & $4.3\pm0.3$   & 10--20\,000   & Miroshnichenko et al. (\cite{Mirosh05}) & yes & no  & Miroshnichenko et al. (\cite{Mirosh05})\\[0.3em]
HD\,87643         & 4.9           & 16\,000       & McGregor et al. (\cite{McGregor}) & yes & ? & McGregor et al. (\cite{McGregor})\\
                  & 5.46          & 16\,200       & Lopes et al. (\cite{Lopes}) & & & \\
                  & 4.0--5.6      & 17--18\,000   & Oudmaijer et al. (\cite{Oudmaijer}) & & & \\[0.3em]
MWC\,84           & $<4.0$        & 18--22\,000   & Miroshnichenko et al. (\cite{Mirosh02}) & & & \\
                  & 5.4           & 18--30\,000   & Hynes et al. (\cite{Hynes}) & & & \\[0.3em]
MWC\,137          & 4.46          & 31\,000       & Hillenbrand et al. (\cite{Hillenbrand}) & & & \\
                  & 5.37          & 30\,000       & Esteban \& Fern\'andez (\cite{Esteban}) & & & \\[0.3em]
MWC\,300          & 5.7--5.85     & 20--22\,000   & Wolf \& Stahl (\cite{WolfStahl})  &  & & \\
                  & $5.1\pm0.1$   & $\sim 20\,000$ & Miroshnichenko et al. (\cite{Mirosh04}) & & & \\[0.3em]
MWC\,314          & $5.95\pm0.2$  & 16\,400--29\,700 & Miroshnichenko et al. (\cite{Mirosh98}) & & & \\[0.3em]
MWC\,349\,A       & $5.7\pm1.0$   & 20--28\,000   & Hofmann et al. (\cite{Hofmann}) & yes & yes & Geballe \& Persson (\cite{Geballe});\\
                  & $6.1\pm0.1$   & 32--35\,000   & Hartmann et al. (\cite{Hartmann}) & &  & Kraus et al. (\cite{Kraus00}) \\[0.3em]
MWC 873           & 6.26          & 13\,000       & Lopes et al. (\cite{Lopes}) & yes & yes & McGregor et al. (\cite{McGregor})\\
                  & $5.0\pm0.4$   & 20\,000       & Miroshnichenko et al. (\cite{Mirosh03}) & & & \\
                  & 6.0           & 11\,500       & Machado et al. (\cite{Machado}) & & & \\[0.3em]
W9 (Ara C)        & 4.86--5.26    & 20--35\,000   & Clark et al. (\cite{Clark}) & & & \\
\hline
\end{tabular}
\label{tab_sg}
\end{table*}

\end{appendix}

                                                                                
\begin{acknowledgements}
                                                                                
I am grateful to Olivier Chesneau, Armando Domiciano de Souza, and Marcelo
Borges Fernandes for exciting discussions on $^{13}$CO band head observations
with VLTI/AMBER, and to the anonymous referee for helpful comments. This
research made use of the NASA Astrophysics Data System (ADS) and of the SIMBAD
database. I gratefully acknowledge financial support from GA\,AV \v{C}R, under
grant number KJB300030701.
                                                                                
\end{acknowledgements}
                                                                                


\begin{thebibliography}{}

\bibitem[2007]{Berthoud}
        Berthoud, M. G., Keller, L. D., Herter, T. L., Richter, M. J., \&
        Whelan, D. G. 2007, ApJ, 660, 461
\bibitem[2004]{BikThi}
        Bik, A., \& Thi, W. F. 2004, A\&A, 427, L\,13
\bibitem[2006]{Bik}
        Bik, A., Kaper, L., \& Waters, L. B. F. M. 2006, A\&A, 455, 561
\bibitem[2007]{Marcelo}
        Borges Fernandes, M., Kraus, M., Lorenz Martins, S., \& de Ara\'ujo,
        F. X. 2007, MNRAS, 377, 1343
\bibitem[1989]{Carr89}
        Carr, J. S. 1989, ApJ, 345, 522
\bibitem[1995]{Carr95}
        Carr, J. S. 1995, Ap\&SS, 224, 25
\bibitem[1993]{Carr93}
        Carr, J. S., Tokunaga, A. T., Najita, J., Shu, F., \& Glassgold, A. E.
        1993, ApJ, 411, L\,37
\bibitem[1993]{Chandler}
        Chandler, C. J., Carlstrom, J. E., Scoville, N. Z., Dent, W. R. F., \&
        Geballe, T. R. 1993, ApJ, 412, L\,71
\bibitem[1996]{Chandra}
        Chandra, S., Maheshwari, V. U., \& Sharma, A. K. 1996,
        A\&AS, 117, 557
\bibitem[1998]{Clark}
        Clark, J. S., Fender, R. P., Waters, L. B. F. M., et al. 1998, MNRAS,
        299, L\,43
\bibitem[2002]{Crowther}
        Crowther, P. A., Hillier, D. J., Evans, C. J., et al. 2002, ApJ, 579,
        774
\bibitem[2004]{Cure04}
        Cur\'{e}, M. 2004, ApJ, 614, 929
\bibitem[2005]{Cure05}
        Cur\'{e}, M., Rial, D.F., \& Cidale, L. 2005, A\&A, 437, 929
\bibitem[2007]{Domiciano}
        Domiciano de Souza, A., Driebe, T., Chesneau, O., et al. 2007, A\&A,
        464, 81
\bibitem[2004]{Evans}
        Evans, C. J., Crowther, P. A., Fullerton, A. W., \& Hillier, D. J.
        2004, ApJ, 610, 1021
\bibitem[1998]{Esteban}
        Esteban, C., \& Fern\'andez, M. 1998, MNRAS, 298, 185
\bibitem[1991]{Farrenq}
        Farrenq, R., Guelachvili, G., Sauval, A. J., Grevesse, N., \&
        Farmer, C. B. 1991, J.~Mol.~Spectrosc., 149, 375
\bibitem[1987]{Geballe}
        Geballe, T. R., \& Persson, S. E. 1987, ApJ, 312, 297
\bibitem[1995]{Gummersbach}
        Gummersbach, C. A., Zickgraf, F.-J., \& Wolf, B. 1995, A\&A 302, 409
\bibitem[1997]{Hanson}
        Hanson, M. M., Howarth, I. D., \& Conti, P. S. 1997, ApJ, 489, 698
\bibitem[1980]{Hartmann}
        Hartmann, L., Jaffe, D., \& Huchra, J. P. 1980, ApJ, 239, 905
\bibitem[1992]{Hillenbrand}
        Hillenbrand, L. A., Strom, S. E., Vrba, F. J., \& Keene, J. 1992,
        ApJ, 397, 613
\bibitem[2003]{Hillier}
        Hillier, D. J., Lanz, T., Heap, S. R., et al. 2003, ApJ, 588, 1039
\bibitem[2002]{Hofmann}
        Hofmann, K.-H., Balega, Y., Ikhsanov, N. R., Miroshnichenko, A. S.,
        \& Weigelt, G. 2002, A\&A, 395, 891
\bibitem[2002]{Hynes}
        Hynes, R. I., Clark, J. S., Barsukova, E. A., et al. 2002, A\&A, 392,
        991
\bibitem[2005]{Rico}
        Ignace, R., \& Gayley, K. G. 2005, The nature and evolution of disks
        around hot stars (San Francisco: ASP), ASP Conf. Ser., 337
\bibitem[2006]{Kastner}
        Kastner, J. H., Buchanan, C. L., Sargent, B., \& Forrest, W. J. 2006,
        ApJ, 638, L\,29
\bibitem[1997]{Diplom}
        Kraus, M. 1997, CO band emission from a rotating disk, Diploma thesis,
        University of Bonn
\bibitem[2000]{PhD}
        Kraus, M. 2000, Modeling of the Near IR Emission from the Peculiar
        B[e] star MWC\,349, PhD thesis, University of Bonn
\bibitem[2005]{KrausBorges}
        Kraus, M., \& Borges Fernandes, M. 2005, in ASP Conf. Ser. 337, The
        nature and evolution of disks around hot stars, ed. R. Ignace \& K. G.
        Gayley (San Francisco: ASP), 254
\bibitem[2006]{KM}
        Kraus, M., \& Miroshnichenko, A. S. 2006, Stars with the B[e]
        Phenomenon, (San Francisco: ASP), ASP Conf. Ser. 355
\bibitem[2007]{KBA}
        Kraus, M., Borges Fernandes, M., \& de Ara\'ujo, F. X. 2007,
        A\&A, 463, 627
\bibitem[2006]{Krausetal06}
        Kraus, M., Borges Fernandes, M., Andrade Pilling, D., \& de Ara\'ujo,
        F. X. 2006, in ASP Conf. Ser. 355, Stars with the B[e] Phenomenon, ed.
        M. Kraus \& A. S. Miroshnichenko (San Francisco: ASP), 163
\bibitem[2008]{Kraus08}
        Kraus, M., Borges Fernandes, M., Kub\'at, J., \& de Ara\'ujo, F. X.
        2008, A\&A, 487, 697
\bibitem[2000]{Kraus00}
        Kraus, M., Kr\"{u}gel, E., Thum, C., \& Geballe, T. R. 2000, A\&A,
        362, 158
\bibitem[1991]{LP}
        Lamers, H.J.G.L.M., \& Pauldrach, A.W.A. 1991, A\&A 244, L5
\bibitem[2001]{Lamers}
        Lamers, H. J. G. L. M., Nota, A., Panagia, N., Smith, L. J., \&
        Langer, N. 2001, ApJ, 551, 764
\bibitem[1998]{Lamers98}
        Lamers, H. J. G. L. M., Zickgraf, F.-J., de Winter, D., Houziaux, L.,
        \& Zorec, J. 1998, A\&A, 340, 117
\bibitem[1992]{Lopes}
        Lopes, D. F., Damineli Neto, A., \& de Freitas Pacheco, J. A. 1992,
        A\&A, 261, 482
\bibitem[2000]{MaMe}
        Maeder, A. \& Meynet, G. 2000, ARA\&A 38, 143
\bibitem[2003]{Machado}
        Machado, M. A. D., \& de Ara\'ujo, F. X. 2003, A\&A, 409, 665
\bibitem[1992]{Magalhaes}
        Magalh\~aes, A. M. 1992, ApJ 398, 286
\bibitem[2006]{Magalhaesetal}
        Magalh\~aes, A. M., Melgarejo, R., Pereyra, A., \& Carciofi, A. C.
        2006, in ASP Conf. Ser. 355, Stars with the B[e] Phenomenon, ed. M.
        Kraus \& A. S. Miroshnichenko (San Francisco: ASP), 147
\bibitem[1988]{McGregor}
        McGregor, P. J., Hyland, A. R., \& Hillier, D. J. 1988, ApJ, 324, 1071
\bibitem[1989]{McGregor89}
        McGregor, P. J., Hyland, A. R., \& McGinn, M. T. 1989, A\&A, 223, 237
\bibitem[2001]{Melgarejo}
        Melgarejo, R., Magalh\~aes, A. M., Carciofi, A. C. \& Rodrigues, C. V.
        2001, A\&A, 377, 581
\bibitem[2003]{MeynetMaeder}
        Meynet, G., \& Maeder, A. 2003, A\&A, 404, 975
\bibitem[2005]{Mirosh05}
        Miroshnichenko, A. S., Bjorkman, K. S., Grosso, M., et al. 2005,
        A\&A, 436, 653
\bibitem[1998]{Mirosh98}
        Miroshnichenko, A. S., Fr\'emat, Y., Houziaux, L., et al. 1998,
        A\&AS, 131, 469
\bibitem[2002]{Mirosh02}
        Miroshnichenko, A. S., Klochkova, V. G., Bjorkman, K. S., \& Panchuk,
        V. E. 2002, A\&A, 390, 627
\bibitem[2003]{Mirosh03}
        Miroshnichenko, A. S., Levato, H., Bjorkman, K. S., \& Grosso, M. 2003,
        A\&A, 406, 673
\bibitem[2004]{Mirosh04}
        Miroshnichenko, A. S., Levato, H., Bjorkman, K. S., et al. 2004,
        A\&A, 417, 731
\bibitem[1996]{Morris}
        Morris, P. W., Eenens, P. R. J., Hanson, M. M., Conti, P. S., \& Blum,
        R. D. 1996, ApJ, 470, 597
\bibitem[1996]{Najita}
        Najita, J., Carr, J. S., Glassgold, A. E., Shu, F., H., \& Tokunaga, A.
        T. 1996, ApJ, 462, 919
\bibitem[1998]{Oudmaijer}
        Oudmaijer, R. D., Proga, D., Drew, J. E., \& de Winter, D. 1998,
        MNRAS, 300, 170
\bibitem[2000]{Pelupessy}
        Pelupessy, I., Lamers, H.J.G.L.M., \& Vink, J.S. 2000, A\&A 359, 695
\bibitem[1997]{Pinsonneault}
        Pinsonneault, M. 1997, ARA\&A, 35, 557
\bibitem[2003]{John}
        Porter, J. M. 2003, A\&A, 398, 631
\bibitem[1992]{Schaller}
        Schaller G., Schaerer D., Meynet G., Maeder A., 1992, A\&AS, 96, 269
\bibitem[1980]{Scoville}
        Scoville, N. Z., Krotkov, R., \& Wang, D. 1980, ApJ, 240, 929
\bibitem[2008]{Searle}
        Searle, S. C., Prinja, R. K., Massa, D., \& Ryans, R. 2008, A\&A, 481,
        777
\bibitem[2008]{Tatulli}
        Tatulli, E., Malbet, F., M\'enard, F., et al. 2008, A\&A, 489, 1151
\bibitem[2005]{Thi}
        Thi, W. F., van Dalen, B., Bik. A., \& Waters, L. B. F. M. 2005,
        A\&A, 430, L\,61
\bibitem[1989]{WinklerWolf}
        Winkler, H., \& Wolf, B. 1989, A\&A, 219, 151
\bibitem[1985]{WolfStahl}
        Wolf, B., \& Stahl, O. 1985, A\&A, 148, 412
\bibitem[2006]{Rens}
        Waters, L. B. F. M. 2006, in ASP Conf. Ser. 355, Stars with the B[e]
        Phe\-no\-menon, ed. M. Kraus \& A. S. Miroshnichenko (San Francisco:
        ASP), 87
\bibitem[1998]{Rens2}
        Waters, L. B. F. M., \& Waelkens, C. 1998, ARA\&A, 36, 233
\bibitem[2000]{Zickgraf00}
        Zickgraf, F.-J. 2000, in ASP Conf. Ser. 214, The Be Phenomenon in
        early-type stars, ed. M. A. Smith, H. F. Henrichs \& J. Fabregat (San
        Francisco: ASP), 26
\bibitem[2006]{Zickgraf06}
        Zickgraf, F.-J. 2006, in ASP Conf. Ser. 355, Stars with the B[e]
        Phenomenon, ed. M. Kraus \& A. S. Miroshnichenko (San Francisco: ASP),
        135
\bibitem[1989]{ZickSchulte}
        Zickgraf, F.-J., \& Schulte-Ladbeck, R. E. 1989, A\&A, 214, 274
\bibitem[1989]{Zick89}
        Zickgraf, F.-J., Wolf, B., Stahl, O., \& Humphreys, R. M.
        1989, A\&A, 220, 206
\bibitem[1986]{Zick86}
        Zickgraf, F.-J., Wolf, B., Stahl, O., Leitherer, C., \& Appenzeller, I.
        1986, A\&A, 163, 119
\bibitem[1985]{Zick85}
        Zickgraf, F.-J., Wolf, B., Stahl, O., Leitherer, C., \& Klare, G. 1985,
        A\&A, 143, 421
\bibitem[1996]{Zick96}
        Zickgraf, F.-J., Humphreys, R. M., Lamers, H. J. G. L. M., et al.
        1996, A\&A, 315, 510
%
\end{thebibliography}
\end{document}